\documentclass[aps,prx,twocolumn,superscriptaddress]{revtex4-2}
\usepackage{amsmath}      % For mathematical symbols and equations
\usepackage{amssymb}      % Additional math symbols
\usepackage{hyperref}     % For hyperlinks in references and citations
\usepackage{float}        % Improved float placement control
\usepackage[table,dvipsnames]{xcolor}
\usepackage{siunitx}      % For SI units, e.g., \si{\meter\per\second}
\usepackage{physics}      % For quantum notation, bra-ket, etc.
\usepackage{bm}           % Bold math for vectors, etc.
\usepackage{mathtools}    % Extension of amsmath
\usepackage{multirow}     % For multi-row cells in tables
\usepackage{dcolumn}      % Align table columns on decimal point
\usepackage{booktabs}
\usepackage{multirow}
\usepackage{adjustbox}
\usepackage[toc]{appendix}
\usepackage[normalem]{ulem}

\begin{document}

\title{Analysis of untrusted-node quantum key distribution from a geostationary satellite}
\author{Thomas Liège}
\email{thomas.liege@lip6.fr}
\affiliation{Sorbonne Université, CNRS, LIP6, F-75005 Paris, France.}
\affiliation{ONERA, DOTA, Paris Saclay University, F-92322 Châtillon, France.}
\author{Perrine Lognoné}
\affiliation{Centre for Advanced Instrumentation (CfAI), Physics Department, Durham University, UK.}
\author{Matteo Schiavon}
\affiliation{Sorbonne Université, CNRS, LIP6, F-75005 Paris, France.}
\author{Caroline B. Lim}
\affiliation{LTE, Observatoire de Paris, Université PSL, Sorbonne Université, Université de Lille, LNE, CNRS, F-75014 Paris, France.}
\author{Jean-Marc Conan}
\affiliation{ONERA, DOTA, Paris Saclay University, F-92322 Châtillon, France.}
\author{Eleni Diamanti}
\affiliation{Sorbonne Université, CNRS, LIP6, F-75005 Paris, France.}
\author{Daniele Dequal}
\affiliation{Telecommunication and Navigation Division, Agenzia Spaziale Italiana, Matera, Italy.}
\affiliation{Connectivity and Secure Communication Directorate, European Space Agency, Noordwijk, Netherlands.}

\date{\today}

\begin{abstract}
    In pursuit of a global quantum key distribution (QKD) network, a service based on untrusted nodes on geostationary satellites could offer wide coverage, continuous operation, and enhanced security compared to the trusted node alternative. Although this scenario has been studied for entanglement-based protocols, such an approach would require large-area telescopes both on the ground and in space. In this work, we analyze the performance of two QKD protocols well adapted to this scenario, namely twin-field (TF) and mode-pairing (MP) QKD, which exhibit high resilience to high-loss channels. Leveraging an in-depth simulation of communication channels corrected with adaptive optics, we assess the expected secret key rates for both protocols in a configuration involving two 50~cm telescopes on board the satellite and ground-based telescopes ranging from 20~cm to 1~m in aperture. Our results show that, in the best case and considering realistic detectors, it is possible to achieve secret key rates on the order of a few hundred bit/s for both TF and MP-QKD. We show, notably, that secret key generation is potentially feasible even with 20~cm ground telescopes, highlighting the high scalability potential of such a configuration. 
\end{abstract}
\begin{figure*}[t]
    \centering
    \includegraphics[width=0.95\textwidth]{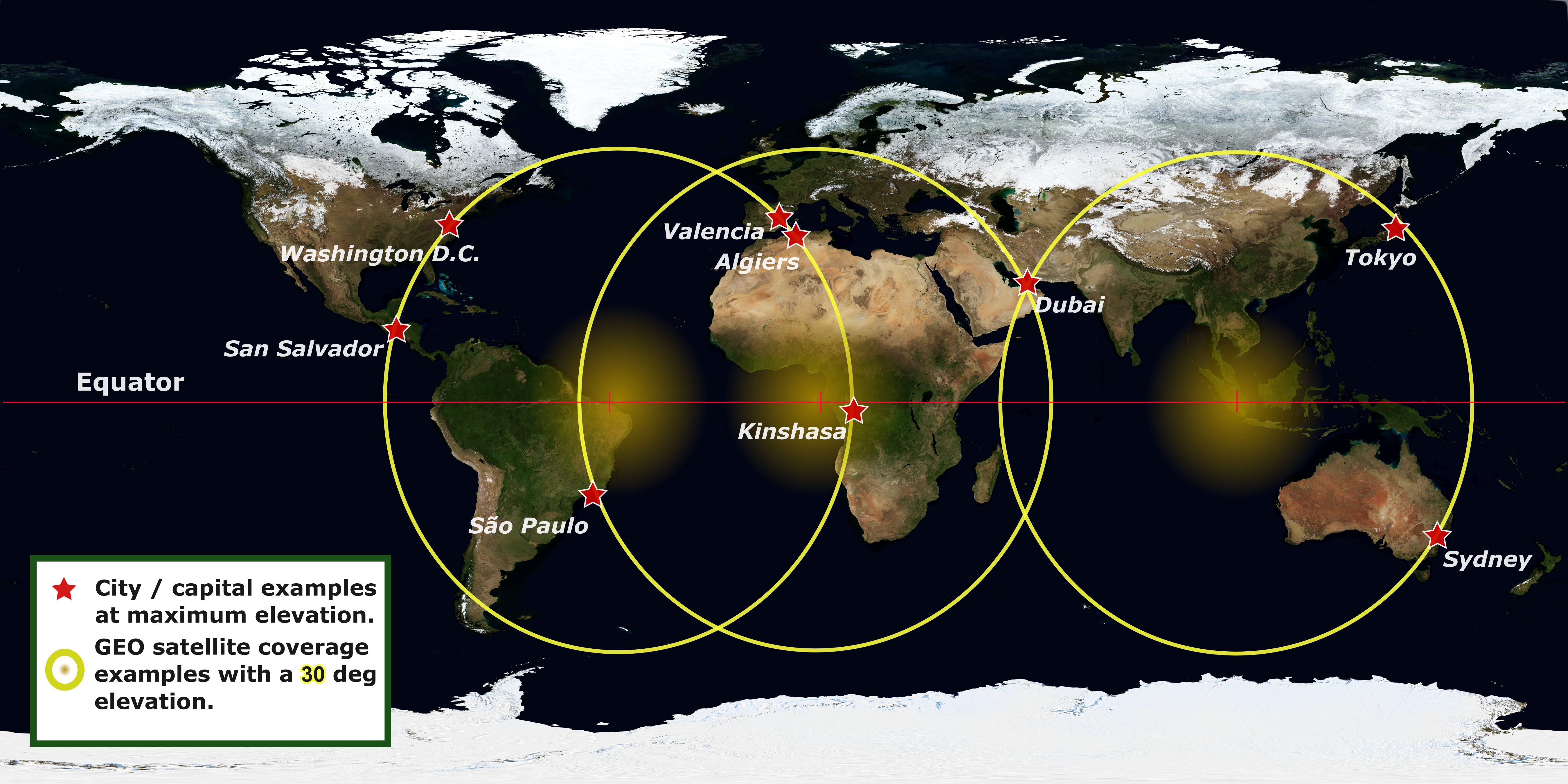}
    \caption{Three examples of coverage of a GEO satellite at 30 degree elevation with different longitudes. A few examples of major cities are given along the coverage paths. The highlighted coverages are approximately at scale. (Map of the Earth: NASA)}
    \label{fig:coverage}
\end{figure*}
\maketitle

% --- MAIN SECTIONS ---
\section{Introduction}

As demands for secure communication increase around the world, satellite-based quantum key distribution (QKD) has emerged as a potential scalable solution to achieve quantum-secured communication on a global scale~\cite{bedington2017progress, liao2017satellite, lee2019updated, dai2020towards}. Current long-distance QKD implementations rely on trusted nodes to relay keys, a solution that can introduce vulnerabilities and jeopardize the security of quantum networks~\cite{salvail2010security, elkouss2013secure, huttner2022long}. A promising solution to address this issue is untrusted-node satellite QKD, which may allow long-distance secret key distribution without requiring trust in intermediary nodes~\cite{fan2021measurement, cao2021hybrid}. 

Geostationary (GEO) satellites are especially valuable in this context. They offer unmatched coverage capabilities, and they allow to continuously serve areas spanning approximately one-third of the planet’s surface. This wide and stable field of view has the potential to permit secret key distribution to multiple ground stations, therefore removing the need for frequent handovers or complex intersatellite relays required for Low-Earth-Orbit satellite constellations~\cite{lee2019updated, gunthner2017quantum}. Such features make GEO satellites particularly attractive for strategic continental and intercontinental links involving, for instance, government data centers, financial hubs, or critical infrastructures, where high-security communication is required. A visual representation of three possible coverages offered by a GEO satellite at different longitudes for an elevation of 30 degrees is given in Fig.~\ref{fig:coverage}. The interest of accessible global-scale links is illustrated in this way, for example, with possible links between North and South America and Europe, between major European cities and Africa, and within a large part of Asia.

A possible solution to perform long-distance QKD via a single untrusted node is based on the distribution of entanglement from a GEO satellite to two optical ground stations (OGS)~\cite{wille2020mission}. The feasibility of this concept has been extensively studied and a critical assessment of the achievable secret key rate has been performed in~\cite{dirks2021geoqkd}, in a configuration with two 0.5~m aperture telescopes on board the satellite and two 2.5~m OGS telescopes. Assuming a 1~GHz pair generation rate, the authors predict a secret key rate of 1.1~bit/s. 

Alternatively, it is possible to remove trust from intermediary nodes using recently introduced QKD protocols, namely twin-field (TF) QKD~\cite{liu2019experimental, wang2018twin, wang2022twin} and mode-pairing (MP) QKD~\cite{zeng2022mode, zhu2023experimental}, which belong to the family of measurement-device-independent (MDI) QKD protocols. Such protocols offer distinct advantages in high-loss scenarios, such as GEO satellite-based communication. Unlike entanglement-based schemes, where successful secret key generation requires both photons of an entangled pair to be detected, TF-QKD requires only a single photon to reach the measurement station, whereas MP-QKD allows for the \textit{a posteriori} pairing of the photons of the pair to be analyzed. This feature significantly improves the resilience to photon losses, resulting in a key rate scaling with the square root of the transmission efficiency, instead of linearly as in the case of entanglement-based protocols. This property has led to the demonstration of the longest ground-based QKD links without the use of intermediary trusted nodes to date~\cite{minder2019experimental, chen2020sending, liu2019experimental}.

In this study, we investigate the performance of the TF-QKD protocol in its so-called sending-or-not-sending version~\cite{wang2018twin} and the MP-QKD protocol \cite{lu2024asymmetric} through GEO satellite channels. We focus on the critical issues affecting such a communication channel, such as atmospheric turbulence, beam divergence, and other transmission losses. To address these issues, we employ advanced simulation methods that incorporate adaptive optics beam pre-compensation, fiber/free-space coupling, and error correction techniques. We assess the performance of TF-QKD and MP-QKD for GEO satellites under realistic conditions, demonstrating their potential as robust, high-performance protocols for untrusted-node satellite QKD over continental and intercontinental distances.

\section{Channel modeling}

Satellite-based MDI-QKD-type protocols, such as TF-QKD and MP-QKD, rely on an uplink exchange between two optical ground stations acting as senders of quantum states, and a satellite acting as receiver (see Fig.~\ref{fig:scheme}). Hence, to evaluate the performance of these protocols, it is necessary to model the losses experienced by the optical beam through its propagation in the atmospheric channel. In this work, we consider an uplink between two optical ground stations~(OGS) and a geostationary~(GEO) satellite pre-compensated by adaptive optics~(AO). We assume that the satellite provides a classical downlink channel, used as a reference measurement beacon for the AO correction computation.

\subsection{Link loss model}
\label{sec:model_sys}
During propagation, the optical beam is affected by different sources of loss, independent of each other. Therefore, the total transmission efficiency can be factorized as follows:
\begin{equation}
    \tau = \eta_{\text{turb}} \eta_{\text{jitter}} \tau_{\text{abs}} \tau_{\text{syst}}\tau_{\text{geom}}.
\end{equation}
Each of these loss factors can be described either as constant or as variable.
Constant losses comprise: internal optical system loss, $\tau_{\text{syst}}$, loss induced by the atmospheric molecular absorption, $\tau_{\text{abs}}$, and geometrical loss, $\tau_{\text{geom}}$, which is induced by the beam divergence~\cite{vedrenne2021performance}. Geometrical loss is a function of the emission~(OGS) and reception~(satellite) telescope aperture diameters, and is expressed as:
\begin{equation}
    \tau_{\text{geom}}= \left(\frac{\pi D_{\text{OGS}} D_{\text{sat}}}{4\lambda L_{\text{OGS-sat}}}\right)^2, 
\end{equation}
where $D_{\text{OGS}}$ is the aperture diameter of the OGS telescope, $D_{\text{sat}}$ is the aperture diameter of the satellite telescope, $\lambda$ is the beam wavelength and $L_{\text{OGS-sat}}$ is the distance between the OGS and the satellite.

Variable losses are induced by the satellite pointing jitter, $\eta_{\text{jitter}}$, and the atmospheric turbulence, $\eta_{\text{turb}}$. The latter term includes the effect of the adaptive optics pre-compensation, considered in this work as a mitigation strategy for reducing the impact of atmospheric turbulence~\cite{bonnefois2022feasibility}.
We model jointly the turbulence effect and the static misalignment of the OGS, as both induce a beam displacement in the satellite plane (constant in the case of the misalignment, and variable in the turbulent case, also known as beam wander).
The statistics of the random variables $\eta_{\text{jitter}}$ and $\eta_{\text{turb}}$ will be described in the following section.

\subsection{Variable loss model and statistics}
\subsubsection{Turbulence effects and beam pre-compensation}
\label{section:turbulence_induced_losses}
\begin{figure}[h]
    \centering
    \includegraphics[width=1\linewidth]{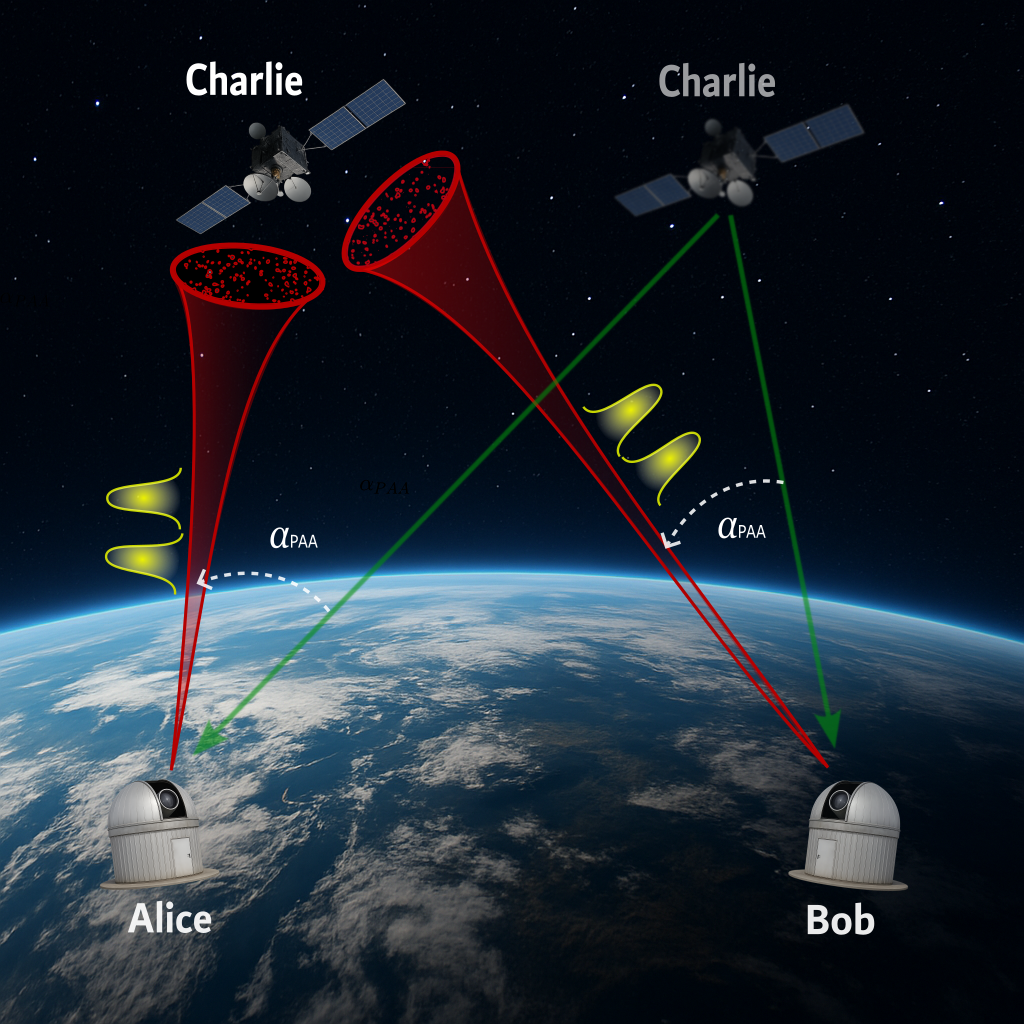}
    \caption{Sketch of the OGS-GEO bidirectional untrusted-node QKD link geometry for a given point-ahead angle, $\alpha_{\text{PAA}}$ (not at scale). The downlink beacon signal optical path is represented in green, the uplink quantum exchange is represented in red. Since the OGS are not located on the equator, the point-ahead angle has to be introduced to account for geometrical variation of the optical path.}
    \label{fig:scheme}
\end{figure}

A crucial element in determining the end-to-end transmission efficiency of a free-space communication system is the divergence of the beam and the spatial fluctuations of the optical pattern in the far-field plane. Although divergence close to the diffraction limit can be achieved by optical telescopes, the distortion of the wavefront introduced by atmospheric turbulence can quickly degrade this ideal value, leading to a wider beam, formations of light speckles and beam wandering, eventually resulting in a reduced transmission performance. A mitigation strategy that can be adopted in this scenario is the use of an AO system to pre-compensate the optical beam to flatten its wavefront after the turbulent layers, with the aim of producing a beam close to the diffraction limit in the satellite plane.
However, due to satellite motion,  a point-ahead angle (PAA) separates the uplink optical path from the downlink, which is used to probe the turbulence and calculate the required pre-compensation. Therefore, as the two beams do not propagate through the same turbulence, the pre-compensation is suboptimal. A visual representation of the problem studied is given in Fig.~\ref{fig:scheme}. Despite being suboptimal, this approach, recently demonstrated on a ground-to-GEO satellite link~\cite{hristovski2024pre,vedrenne2024first}, has been shown to largely improve, from 10 to more than 20~dB, the mean value and also the stability of the flux received at the satellite.

\begin{figure}
    \centering
    \includegraphics[width=0.9\linewidth]{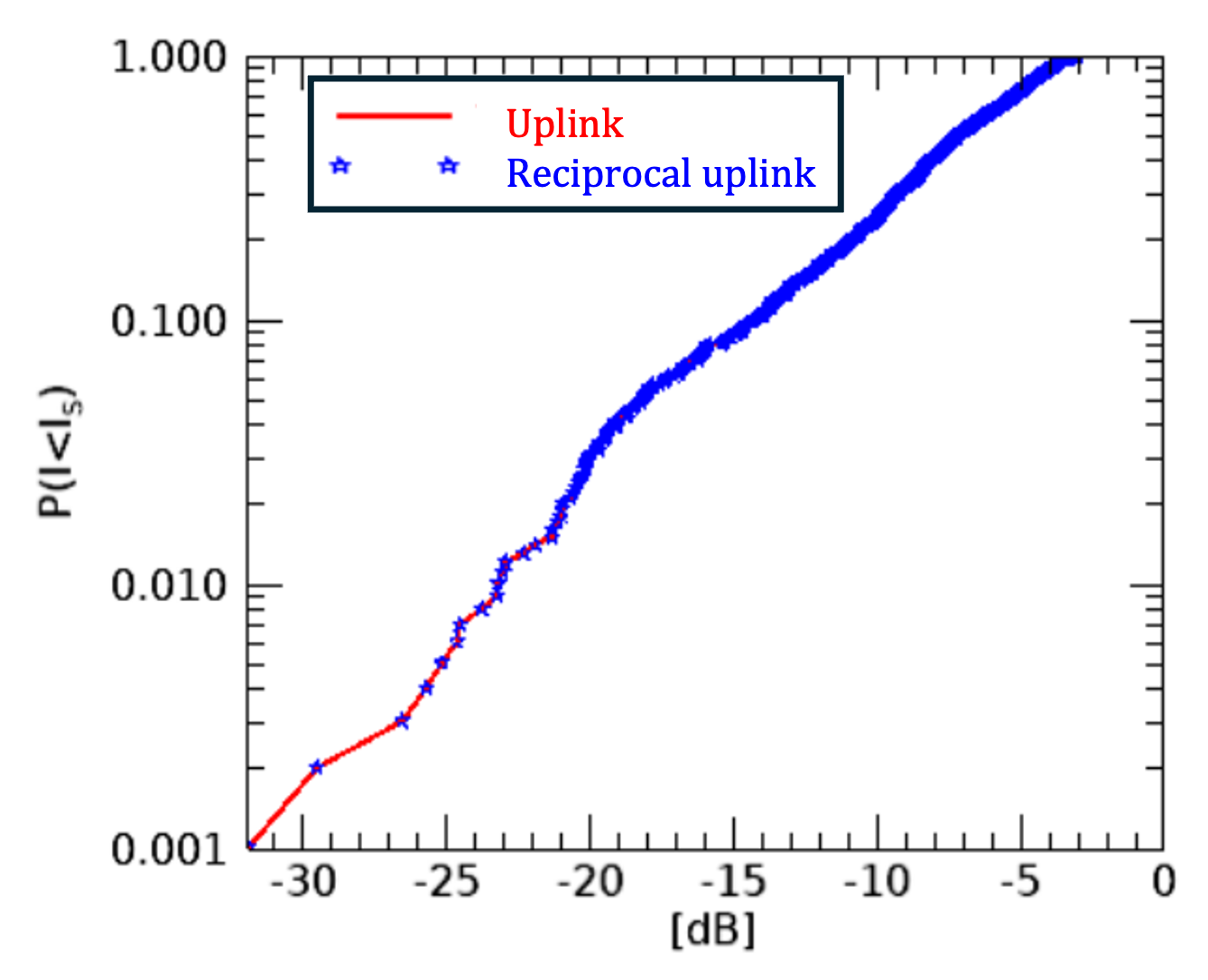}
    \caption{Cumulative density function of the coupling efficiency of a pre-compensated uplink coupled to the satellite compared to the flux of the reciprocal (\emph{i.e.}, downlink) corrected by adaptive optics, coupled to a Gaussian emission  mode, simulated using numerical wave optics simulation tools (from ONERA Pilot model~\cite{lognone2023optimization}).}
    \label{Fig:Figure_reciprocity}
\end{figure}

To model the pre-compensated uplink losses induced by atmospheric turbulence, we use the reciprocity principle. This principle has been first analytically studied in~\cite{shapiro2012reciprocity}, experimentally demonstrated in~\cite{parenti2012experimental,yao2019analysis,lognone2022new} and exploited to simulate pre-compensated ground-to-satellite links in~\cite{PhysRevA.93.033860,Farley:22,lognone2023phase}, as illustrated in figure~\ref{Fig:Figure_reciprocity}. The main advantage of this principle is to allow using plane wave downlink analytical and numerical simulation frameworks, extensively developed for astronomy. 

The principle states that the coupling of the uplink turbulent mode to the satellite receiver mode (in the satellite plane), is equal to the coupling of the satellite receiver mode back-propagated towards the OGS to the transmitter emission mode.
It allows to model the pre-compensated uplink losses as the losses of a downlink considering a deviation of $\alpha_{\text{PAA}}$ between the beam used to probe the turbulent-distorted wavefront and the beam to be corrected via AO. The complete model of $\eta_{turb}$ using a pseudo-analytical approach is derived in Appendix~\ref{appendix:atm_turb}.

For this analysis, we consider two types of AO correction. First, we consider the state of the art~(SoA) correction that consists in applying the on-axis downlink phase correction to the off-axis uplink. In this case, the PAA angular shift between uplink and downlink will lead to phase residuals from the on-axis/off-axis phase mismatch. Second, we consider an advanced pre-compensation method, relying on a minimum mean square error (MMSE) estimation, called here MMSE. This MMSE method relies on the estimation of the phase at PAA based on the on-axis downlink phase and amplitude measurements, and was shown to greatly reduce the SoA phase residuals and the turbulence induced coupling losses~\cite{lognone2023phase}.

Furthermore, we model the static pointing error from the OGS by adding a constant misalignment phase tip-tilt to the simulated phase. Indeed, from the reciprocal point of view, the pointing error on sky is equivalent to a tilt (or tip) in the OGS aperture plane. We model the corresponding static phase term as $\Phi_{\text{misp}}(\mathbf{r})=a_2 Z_2(\mathbf{r})$, where $a_2 = \pi D \Delta \alpha/2\lambda$, $\Delta \alpha$ is the  OGS misalignment error, and $Z_2(\mathbf{r})$ is the second Zernike mode (tip). Finally, given the phase statistics, phase vectors are generated and 2D numerical samples of the phase and complex field are synthesized. These are then numerically coupled to the Gaussian mode to obtain the phase contribution to the coupling. We provide complete model steps and formulas in Appendix~\ref{appendix:turb_model}.

\subsubsection{Satellite jitter model}
Next, we also model the fluctuating losses caused by the satellite pointing jitter by applying the reciprocity principle. This allows us to use tools from the literature~\cite{vasylyev2012toward,acosta2024increasing} that apply to downlink scenarios.

In a downlink scenario, the satellite jitter induces a random beam displacement around the ground station telescope aperture. The probability distribution of the deflection distance $r$, that is, the distance between the optical beam center and the center of the aperture, is expressed as: 
\begin{equation}
\label{Weibull}
    P(r) = \frac{r}{\sigma_r^2}\exp(-\left(\frac{r}{\sqrt{2}\sigma_r}\right)^2),
\end{equation}
which corresponds to a Weibull probability distribution with zero mean and a standard deviation dependent on $\sigma_r \simeq L_{\text{OGS-sat}}\theta_{\text{jitter}}$, where $L_{\text{OGS-sat}}$ is the propagation distance and $\theta_{\text{jitter}}$ the satellite jitter angle.

Then, knowing the probability distribution of the deflection distance, the transmission efficiency corresponding to each distance from the center of the aperture can be calculated as:
\begin{equation}
    \eta_{\text{jitter}} = \eta_0\exp(-\left(\frac{r}{\beta}\right)^\alpha),
\end{equation}
where $\eta_0$ is the maximal transmission efficiency, and $\alpha$ and $\beta $ are the shape and scale parameters, whose description can be found in~\cite{vasylyev2012toward}. We compute $\eta_{\text{jitter}}$ through numerical simulations where we perform a run of 10000 random occurrences of the variable $r$, according to Eq.~(\ref{Weibull}).

\subsection{End-to-end channel simulation}
\label{sec:PDTE}
As a final step, we calculate the complete probability distribution of the total transmission efficiency $\tau$ by considering all the effects described in the previous sections. The resulting probability distribution of the transmission efficiency $\tau$ of the quantum channel is given by~\cite{acosta2024analysis}:
\begin{equation}
    \text{PDTE}(\tau) = 
    \int_{-\infty}^{\infty} P_{\text{AO}}(x) P_{\text{jitter}} \left( \frac{\tau}{x} \right) \frac{1}{|x|}  dx,
\end{equation}
where $P_{\text{jitter}}$(x) and $P_{\text{AO}}$(x) describe, respectively, the probability distribution of the transmission efficiency as a function of the satellite jitter, $\eta_{\text{jitter}}$, and of the atmospheric turbulence combined with the OGS static pointing error, $\eta_{\text{turb}}$.

Finally, by including the fixed loss terms, we derive the end-to-end channel transmittance in the GEO exchange.

\section{Secret key rate estimation}

In this work, we study two different QKD protocols that belong to the MDI-QKD family: Twin-Field QKD and Mode-Pairing QKD. Both protocols theoretically surpass the so-called PLOB repeaterless bound~\cite{zhong2019proof, zhong2021proof, li2024field, zhang2025experimental}, and feature a key rate scaling proportional to the square root of the total channel attenuation $\tau$. To simplify the analysis, we assume the two optical links involved in our scenario to have the same characteristics, \emph{i.e.}, they are modeled with the same PDTE. It is worth underlining that this is not a limiting choice, as an extension of the analysis to different configurations (for instance, OGS with different aperture size or distance from the satellite) can be addressed by changing the intensity of the transmitted pulses. As demonstrated in~\cite{zhou2019asymmetric, lu2024asymmetric}, static channel asymmetries can be pre-compensated, achieving a key rate similar to the symmetric case. Although this technique can be used for predictable fixed losses (like the geometrical one), it cannot be used for fluctuating effects, such as pointing errors or turbulence effects. In this case, it is still possible to perform a symmetrization of the channels by probing the instantaneous transmission efficiency with a beacon laser and adding losses to the channel with the highest transmission efficiency. In the following, this case is referred to as the ``compensated'' case, while the option to leave the asymmetry of the two channels is referred to as the ``non-compensated'' case.

\subsection{Twin-field QKD}
The twin-field (TF) QKD protocol, proposed in~\cite{lucamarini2018overcoming}, can be understood as a derivation of prepare-and-measure QKD with phase encoding. By generating the two pulses at two different locations and looking at the phase relation between the two, it is possible to double the distance covered by the protocol with respect to the prepare-and-measure version. Although this feature greatly improves the achievable distance, it also comes at the cost of having to stabilize the optical phase of the two pulses. This has been achieved on ground~\cite{liu2019experimental,liu2023experimental}, and more recently over a free-space link~\cite{li2025free}, but the extension to space will represent a significant challenge due to satellite motion, atmospheric effects and long distance between terminals. In the scenario considered in this work, the satellite motion can be minimized due to the use of a GEO satellite, while atmospheric effects are analyzed in detail and compared in the following to fiber-based experimental demonstrations. As in~\cite{takenaka2017satellite, yin2017satellite}, we consider here the implementation of an active feedback correction based on a beacon laser.

The asymmetric sending-or-not-sending TF-QKD protocol allows the use of the standard protocol while tolerating channel asymmetries. As demonstrated in~\cite{wang2020simple}, channel asymmetries have a serious impact on the performance of the protocol, and being able to compensate for the asymmetries can lead to huge improvements. As explained above, the compensated case is simulated by increasing the attenuation of one of the channels on board the satellite (Charlie) (see Fig.~\ref{fig:scheme}) so that both links have the same transmittance. This increases the overall attenuation, but allows us to consider asymmetric channels while having symmetric attenuation profiles. The security of the protocol is described in Appendix~\ref{sec:TF_security_model}. 

Its performance is evaluated based on key metrics, including the $X$-basis bit error rate, $e_{X}$, the $Z$-basis bit error rate, $e_{Z}$, and the overall secret key rate, $R$.

\subsection{Mode-pairing QKD}
\label{sec:mp_qkd}
Mode-pairing (MP) QKD, proposed in~\cite{zeng2022mode}, can be seen as a derivation of time-bin encoding, where the photons of the time-bin pair are selected \textit{a posteriori} based on the event of a photon detection.
As for TF-QKD, MP-QKD key rate scales as the square root of the transmission efficiency, but by relying on frequency locking instead of global phase locking. This makes the MP-QKD protocol more practical for real-world applications. We consider again asymmetric channels, and we use the model given in~\cite{lu2024asymmetric} and described in Appendix~\ref{sec:MP_security_model}. In this model, Alice and Bob send weak coherent pulses to Charlie. Then, Charlie performs an interference measurement and publicly announces the outcomes. Alice and Bob then pair the detected pulses while making sure that the interval between the paired pulses does not exceed the so-called maximal pairing length $L_{\text{max}}$, which limits the quantum bit error rate (QBER) introduced by the phase drift between the matched pairs. Paired pulses are assigned either to the \(Z\)- or \(X\)-basis based on intensity criteria. After parameter estimation, Alice and Bob use a decoy-state analysis to bound the key parameters, enabling the extraction of a secret key through error correction and privacy amplification.

\subsection{Impact of detector quality and propagation phase fluctuation}
\label{sec_detector}
In the following analysis, we will consider an attenuation from $100$~dB to $130$~dB. Since the key rate drops in a region that is correlated to the dark count rate of the single-photon detectors, the choice of the detectors is important for the simulation of the OGS-GEO QKD exchange. 
 
Commercial superconducting nanowire single photon detectors (SNSPD) for ground applications provide an efficiency up to $\eta_D = 90\%$, and a dark count rate of a few Hz~\cite{zhu2023experimental, lu2024asymmetric, zeng2022mode, li2024field, zhu2024field, chen2024twin}.  The deployment of SNSPD technology in space is still an area of research and development, with limited results so far. These include notable breakthroughs achieving a detection efficiency of $\eta_D \sim$ 50\% for a dark count rate of $Y_0 = 100$~Hz with a FWHM time jitter of 48~ps~\cite{you2018superconducting}. In our analysis, we consider an overall system jitter of 100~ps, which includes the detector jitter and results in a detection window of 400~ps. In Section~\ref{sec:comparison_snspd}, we compare the secret key rate performance for three detection scenarios: an optimistic case, with dark count rate $Y_0 = 25$~Hz $\Leftrightarrow$ dark count probability $p_d = 10^{-8}$ and detection efficiency $\eta_D = 70\%$; a pessimistic case with the parameters demonstrated in~\cite{you2018superconducting}, \emph{i.e.}, $Y_0 = 100$~Hz $\Leftrightarrow p_d = 4 \times 10^{-8}, \eta_D = 50 \%$; and an idealized case of state-of-the-art commercial ground detectors brought to space, with $Y_0 = 1$~Hz $\Leftrightarrow p_d = 4 \times 10^{-10}, \eta_D = 90 \%$. 

A second effect that needs to be considered is the phase mismatch between the two links and its evolution. Indeed, this mismatch plays a key role in the calculation of the misalignment error, $e_d$, impacting the overall key rate. In these QKD protocols, the misalignment error represents the interferometric error that arises due to imperfect phase or polarization matching between the signals sent by Alice and Bob on Charlie's side. Thus, this error is directly linked to the angles $\theta$ (phase) and $\phi$ (polarization) in the TF-QKD protocol; see Appendix~\ref{sec:TF_security_model}. A specific model to estimate $e_d$ is given for the case of MP-QKD in Appendix~\ref{sec:phase_fluct} and its impact on the MP-QKD performance is described in Appendix~\ref{sec:MP_security_model}.

\begin{figure}[h]
    \centering
    \includegraphics[width=1\linewidth]{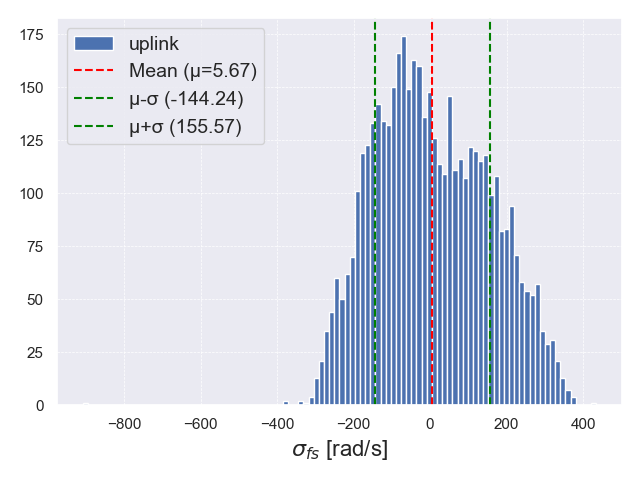}
    \caption{Probability density function (PDF) of the free-space phase drift taken from the VERTIGO dataset. The mean value is highlighted in red, the standard deviation of the distribution is highlighted in green. Both are respectively approximated as $\mu = 0$ rad/s, and $\sigma_{\text{fs}} = 150$ rad/s for the simulations.}
    \label{fig:histo}
\end{figure}

Although the MP-QKD protocol does not require phase matching, it requires the phase difference to remain constant between the two pulses of the matched pair. A time evolution of the phase, due to source drifts or the transmission channel, would increase the QBER of the $X$-basis. To limit this effect, it is possible to set a maximal pairing length, $L_{\text{max}}$, to allow for pulse pairing. 
Increasing $L_{\text{max}}$ allows to consider more pairs in post-processing but would also increase the $X$-basis error rate due to phase fluctuations. To estimate the time evolution of the phase between the matched pulses, three factors need to be considered: the phase error due to the linewidth of the laser, the phase error due to the frequency offset between Alice and Bob, and the phase drift due to free-space propagation. The first two effects depend on the lasers used. The third factor depends on the propagation of the beam through atmospheric turbulence. Most of the literature on MDI-QKD considers the phase drift introduced by a fixed length fiber, while in this work we consider the phase drift due to propagation in the case of an OGS-GEO uplink exchange with an elevation of $\theta_{\text{elev}} = 30$~deg. We estimate the phase drift with the \textit{Very High Throughput Satellite–Ground
Optical Feeder Link} (VERTIGO) simulation tool developed by ONERA, which has been described in~\cite{le2021h2020}. The distribution of the phase drift for the considered free space channel using a severe turbulence condition (MOSPAR 90-90 turbulence profile) is given in Fig.~\ref{fig:histo}.

For simplicity and in order to fit in the phase model proposed in~\cite{zhou2025optimization}, we consider that the free-space drift of the phase follows a Gaussian distribution with zero mean and standard deviation $\sigma_{\text{fs}} = 150$~rad/s for each uplink channel. In comparison, the phase drift standard deviation due to fiber propagation ranges from 6~rad/ms~\cite{lucamarini2018overcoming} to 20~rad/ms~\cite{zhu2023experimental}.

Looking now at TF-QKD, this protocol requires phase locking, but the model for the phase evolution during the free-space propagation remains the same. As the phase drift rate is much smaller than in fiber-based links, we believe it would be possible to keep the phase mismatch within the range of $\theta <$ 0.5~rad, as demonstrated experimentally in~\cite{li2025free}. In this case, the impact of phase drift on the residual phase error would be $e_{d}^{\theta} = \sin(\frac{\theta}{2})^2 = 0.1\%$ contributing to the total misalignment error in the QBER. We note that, given the free-space phase drift standard deviation of 150~rad/s, a feedback loop correcting every millisecond would suffice to compensate for the phase drift to the required level. 
Finally, regarding possible polarization mismatch, we consider the presence of polarization beam splitters at Charlie's side and polarization controllers at Alice's and Bob's sides. With this strategy, we estimate a residual polarization error of $e_d^P = \sin(\frac{\phi}{2}) \sim 0.1\%$.
\section{Simulation and discussion}
\label{sec:simulation}
In the following, we simulate the performance of TF-QKD and MP-QKD using our atmospheric channel model. We provide, in particular, the plots of the coupling efficiency, PDTE and secret key rate for five different OGS aperture diameters $D_{\text{OGS}}$, ranging from 20~cm to 100~cm. The diameter of the telescope aperture on board the satellite is $D_{\text{sat}} = 50$~cm, and the satellite is assumed to be located in the GEO stationary orbit $h_{\text{sat}} = 35786$~km at an elevation of $\theta_{\text{elev}} = 30$~deg, with respect to both OGSs. This results in an OGS-to-satellite distance of $L_{\text{OGS-sat}} = 38608.88$~km. In this configuration, the point-ahead angle is $\alpha_{\text{PAA}} = 18.5~\mu$rad \cite{lognone2023phase,mengali2020ground}, and we consider an OGS misalignment error of $\Delta \alpha = 0.2~\mu$rad~\cite{vedrenne2021performance}. An overview of all the parameters is given in Table~\ref{tab:param}. We recall that $\tau_{\text{syst}}$ is the fixed attenuation accounting for optical system losses, while $\theta_{\text{jitter}}$ is the residual tracking error from the satellite, $F$ is the repetition rate, and $f_{\text{EC}}$ is the error correction efficiency used in the QKD protocol. Finally, $\sigma_\nu$ and $\Delta \nu$ are respectively the frequency uncertainty standard deviation caused by the linewidth of the laser and the estimated frequency difference between Alice and Bob transmitters.

\begin{table}[h]
    \centering
    \begin{tabular}{cccccccc}
        \toprule
        $\alpha_{\text{PAA}}$ & $\Delta \alpha$ & $D_{\text{sat}}$ & $\theta_{\text{elev}}$ &  $\tau_{\text{\text{syst}}}$ & $\theta_{\text{jitter}}$\\
        \midrule
        18.5 $\mu$rad & 0.2 $\mu$rad & 50 cm & 30 deg & 2.8 dB & 0.07 $\mu$rad\\
        \midrule
        $p_d$ & $\eta_{D}$ & $F$ & $\sigma_\nu$ & $\Delta \nu$  & $f_{\text{EC}}$\\
        \midrule
        $10^{-8}$ & 70 \% & 2.5 GHz & 1 kHz & 0.1 kHz & 1.1  \\
        \bottomrule
    \end{tabular}
    \caption{Values of the parameters used in our simulations.}
    \label{tab:param}
\end{table}

\subsection{Atmospheric turbulence modeling}
The turbulence profiles used to simulate the atmospheric channel are taken from the MOSPAR database, constructed from astronomical site measurements (Paranal for upper layers, and Tenerife for the lower layers, linked using a Monin–Obhukov similitude law to account for day or nighttime)~\cite{osborn_10.1093/mnras/sty1070,sprung2013characterization,acosta2024analysis}. Using these databases containing more than 10000 measurements, the MOSPAR profiles are then constructed to be statistically representative of the turbulent integrated parameters $r_0$ and $\theta_0$, which describe the atmospheric conditions. 

We consider pessimistic atmospheric conditions at nighttime, meaning that only $25$\% of the time the turbulence conditions are worse than the ones from the dataset. In this turbulence scenario, the integrated parameters are: $r_0=25$~cm for the Fried parameter describing the total turbulence strength at $30$~degree elevation and at 1550~nm, $\theta_0=8.51$~$\mu$rad for the isoplanatic angle referring to the angular decorrelation of the turbulence, and $\sigma^2_\chi=0.03$ for the log-amplitude variance giving the scintillation conditions. These parameters are shown to be consistent with recent measurements in an urban environment~\cite{beesley2025demonstration}. 
The number of adaptive optics corrected modes is tuned to keep the phase fitting error roughly constant with an increasing aperture diameter~\cite{lognone2023optimization}. This fitting error corresponds to the phase uncorrected by the AO system and its variance is chosen to be equal to $\sigma^2_{\Phi_{\text{fit}}}=0.01$~rad$^2$. The number of modes corrected for each aperture diameter can be found in Table~\ref{tab:modes}.

\begin{table}[h]
\centering
\setlength{\tabcolsep}{8pt}
\renewcommand{\arraystretch}{1.3}
\begin{tabular}{ccc cc}
    \toprule
    \multirow{2}{*}{$D_{\text{OGS}}$ (cm)} & \multirow{2}{*}{$N_{\text{corr}}$} &
    \multicolumn{2}{c}{$\eta_{\text{turb}}$ (mean $\pm$ standard deviation)} \\
    \cmidrule(lr){3-4}
    & &MMSE & SoA Correction \\
    \midrule
    20  & 45 &  $0.73 \pm 0.1$  & $0.72 \pm 0.18$ \\
    40  & 91 & $0.66 \pm 0.12$ & $0.62 \pm 0.14$ \\
    60  & 136&  $0.61 \pm 0.11$ & $0.53 \pm 0.15$ \\
    80  & 190&  $0.58 \pm 0.11$ & $0.45 \pm 0.15$ \\
    100 & 231&  $0.56 \pm 0.1$  & $0.40 \pm 0.15$ \\
    \bottomrule
\end{tabular}
\caption{Number of AO correction modes and correction efficiencies for each OGS aperture diameter.}
\label{tab:modes}
\end{table}

\begin{figure}[h]
    \centering
    \includegraphics[width=1\linewidth]{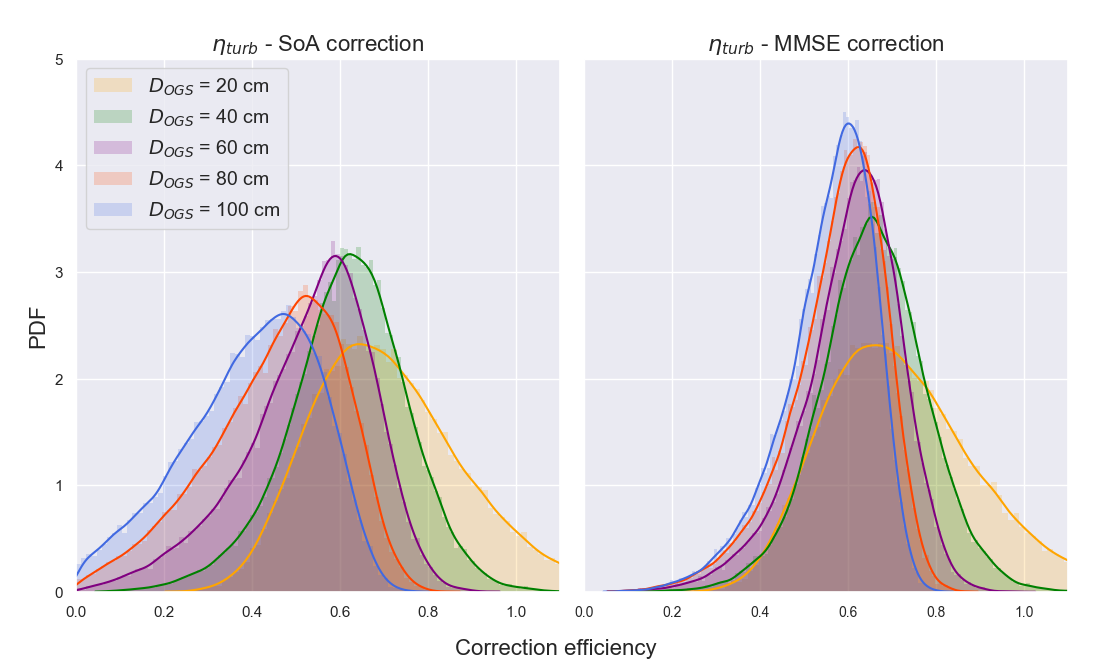}
    \caption{PDF of $\eta_{\text{turb}}$ with respect to a flat wavefront. The state-of-the-art and  MMSE corrections are shown, respectively, on the left and right of the figure.}
    \label{fig:coupling_MMSE}
\end{figure}

The PDF of the turbulence correction efficiencies for each channel with the SoA correction and with the MMSE estimator are given in Fig.~\ref{fig:coupling_MMSE} for several OGS aperture diameters. We note that, as the PDF quantify the ratio between the AO-corrected wavefront and an ideal flat wavefront, in some cases it is possible to have a concentration of the beam within the receiving aperture, thus explaining the occurrences of PDF above 1. The numerical comparison between the two correction schemes is given in Table~\ref{tab:modes}. 

We observe an increasing impact of the MMSE estimator on the quality of the coupling efficiency as the OGS aperture diameter increases. This can be explained by two factors. First, the MMSE estimator is a phase estimator and is therefore more efficient when the phase contribution to the coupling is dominant. Although small aperture turbulent losses are dominated by the amplitude contribution (and only feature a small phase contribution to the coupling fluctuations), large aperture scenarios are dominated by the phase contribution to the coupling~\cite{conan2019adaptive}. This is shown in Appendix~\ref{appendix:atm_turb}. Second, larger apertures capture a larger amount of phase and amplitude, and, therefore, the correlations between the on-axis measurements and the phase at PAA are stronger, which benefits the MMSE estimation.

\subsection{Probability density of transmission efficiency}
\label{sec:PDTE_simu}
\begin{figure}[h]
    \centering
    \includegraphics[width=1\linewidth]{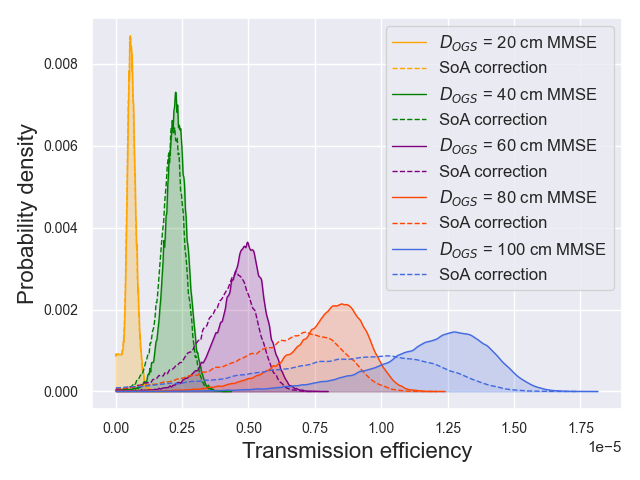}
    \caption{PDTE comparison for one channel with AO pre-compensation, considering the MMSE or the SoA correction, for different OGS aperture diameters.}
    \label{fig:pdte}
\end{figure}

We can now calculate the final PDTE taking into account the turbulent losses, $\eta_{\text{turb}}$, the satellite jitter losses, $\eta_{\text{jitter}}$, the absorption and scattering losses, $\tau_{\text{abs}}$ — obtained thanks to the $\text{MODTRAN}^{\textregistered}$ tool \cite{berk2014modtran} — and the fixed attenuation, $\tau_{\text{syst}} = 2.8$~dB accounting for optical system losses. The residual tracking error from the satellite is $\theta_{\text{jitter}} = 0.07~\mu$rad~\cite{cantore2024link}. The results are shown in Fig.~\ref{fig:pdte}. Overall, the average attenuation reached for each channel ranges from 50~dB to 65~dB, depending on the OGS aperture diameter. The main effect that contributes to the total attenuation is the geometrical loss as it dramatically increases the probability of having lower values of transmission efficiency.

\subsection{Secret key rate estimation}
\label{sec:mdi}
We assess the performance of TF-QKD and MP-QKD first by varying the average intensity per pulse, $\mu$, set to be the same for Alice and Bob, and then as a function of the OGS aperture diameter, by taking the optimal $\mu$.

\subsubsection{Twin-field QKD}
\begin{figure}[h]
    \centering
    \includegraphics[width=1\linewidth]{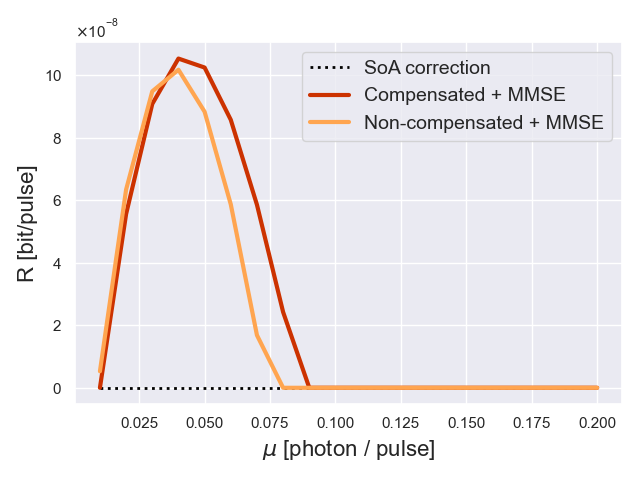}
    \caption{TF-QKD secret key rate performance for $D_{\text{OGS}} = 100$~cm. Other parameters are $p_d = 10^{-8}$, $\eta_D = 70 \%$.}
    \label{fig:TF_60}
\end{figure}

The evolution of the performance of TF-QKD, with respect to the average intensity used per pulse, is shown in Fig.~\ref{fig:TF_60} for $D_{\text{OGS}} = 100$~cm. In this scenario, for the best (compensated + MMSE) case, the secret key rate reaches a maximum value of $R_{\text{max}} \simeq 1.05 \times10^{-7}$~bit/pulse allowing to transmit up to $\sim 260$~bit/s, for $\mu = 0.04$~photon/pulse, with error rates $e_{X} \simeq 1.4 \%$ and $e_{Z} \simeq 22 \%$. The $Z$-basis error rate is quite high but similar to the one obtained in previous experimental results in high attenuation scenarios~\cite{li2025free, chen2024twin} (without actively-odd-parity pairing method, see~\cite{xu2020sending}).

\begin{figure}[h]
    \centering
    \includegraphics[width=1\linewidth]{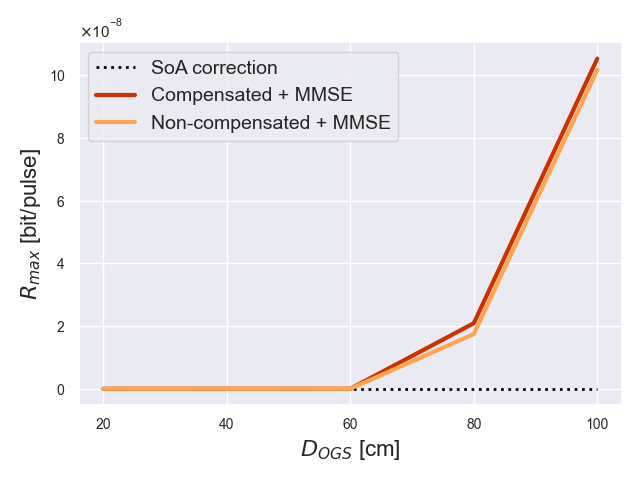}
    \caption{TF-QKD maximal secret key rate reached for each OGS aperture diameter.}
    \label{fig:TF_R}
\end{figure}

By doing the same analysis for every aperture diameter, we obtain the total evolution of the maximum key rate reached in Fig.~\ref{fig:TF_R}. The effect of the channel asymmetry is partially mitigated by the use of the MMSE correction because the asymmetry between the two channels is purely due to the probabilistic nature of fluctuating losses: since the MMSE method decreases the standard deviation of the attenuation distribution, the asymmetry will be less intense for these scenarios. When simulating the SoA correction, without MMSE, we did not obtain a positive key rate for any aperture diameter under these conditions.

\subsubsection{Mode-pairing QKD}

\begin{figure}[h]
    \centering
    \includegraphics[width=1\linewidth]{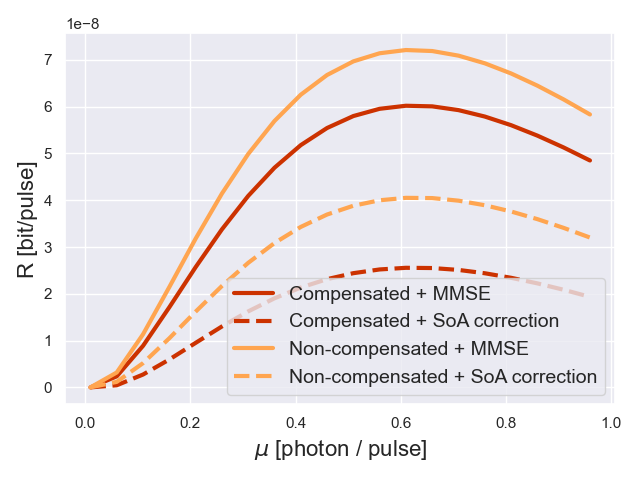}
    \caption{MP-QKD secret key rate performance for $D_{\text{OGS}} = 100$~cm. Other parameters are $p_d = 10^{-8}$, $\eta_D = 70 \%$, $L_{\text{min}} = 100$. $L_{\text{min}}$ is the minimal pairing length introduced to account for the detector dead time; see Appendix~\ref{sec:MP_security_model}.}
    \label{fig:MP_60}
\end{figure}
For MP-QKD, it is also necessary to optimize the maximal pairing length $L_{\text{max}}$. For each OGS aperture diameter, we scan the best key rate reached for $L_{\text{max}} \in [10^3, 10^6]$ and then select the best maximal pairing length for each scenario; more details can be found in Appendix~\ref{sec:qkd_simu}. The work presented in~\cite{zhou2025optimization} gives a similar method to obtain the optimal $L_{\text{max}}$. The performance for $D_{\text{OGS}} = 100$~cm at varying source intensities can be found in Fig.~\ref{fig:MP_60}. For this aperture diameter, in the best (non-compensated + MMSE) case, the secret key rate reaches a maximum value of $R_{\text{max}} \simeq 7.2 \times10^{-8}$~bit/pulse allowing to transmit up to $\sim 180$~bit/s, for $\mu = 0.6$~photon/pulse, with error rates $e_{X} \simeq 2.3 \%$ and $e_{Z} \simeq 0.55 \%$.

\begin{figure}[h]
    \centering
    \includegraphics[width=0.45\textwidth]{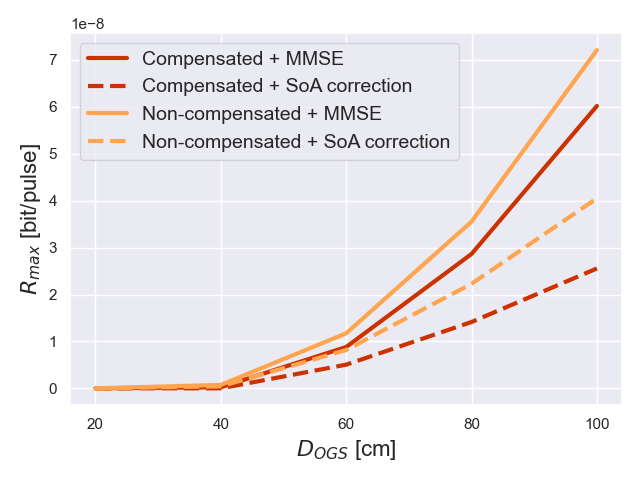}
    \caption{MP-QKD maximal secret key rate reached for each OGS aperture diameter.}
    \label{fig:MP_R}
\end{figure}

By doing the same analysis for every aperture diameter, we obtain the total evolution of the maximum key rate reached in Fig.~\ref{fig:MP_R}.
\label{sec:mp_simu}

We remark that for MP-QKD, contrary to TF-QKD, the results for the non-compensated case are better than for the compensated one. This can be explained considering the impact of the phase difference between two paired pulses: in the compensated case, the overall attenuation is higher, thus resulting in a lower detection probability for each pulse sent. Therefore, the time between two successful paired pulses increases, leading to a higher phase difference, \emph{i.e.}, a higher $X$-basis error rate.

Furthermore, the maximum key rate reached for MP-QKD is approximately of the same order of magnitude as for TF-QKD ($\sim 10^{-7}$). This highlights how well this protocol that does not require global phase locking performs, achieving a secret key rate comparable to that of TF-QKD while being more suitable for practical implementation. 

\subsubsection{Comparison with different SNSPD scenarios}
\label{sec:comparison_snspd}
\begin{figure}[h]
    \centering
    \includegraphics[width=0.45\textwidth]{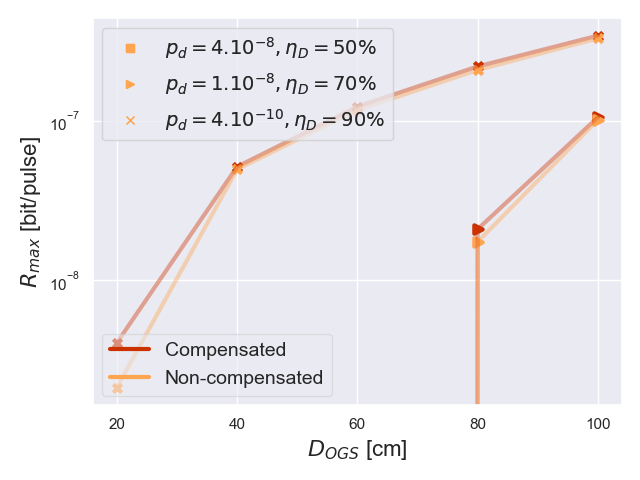}
    \caption{TF-QKD maximal key rate evolution as a function of $D_{\text{OGS}}$ with the MMSE estimator for different detection scenarios (logscale).}
    \label{fig:TF_detection}
\end{figure}

\begin{figure}[h]
    \centering
    \includegraphics[width=0.45\textwidth]{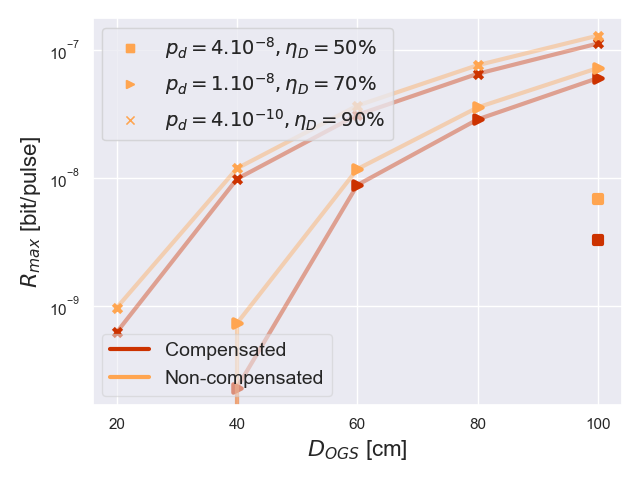}
    \caption{MP-QKD maximal key rate evolution as a function of $D_{\text{OGS}}$ with the MMSE estimator for different detection scenarios (logscale).}
    \label{fig:MP_detection}
\end{figure}

We further analyze the secret key rate performance as a function of the single-photon detector parameters using the values discussed in Section~\ref{sec_detector}.  An overall comparison for TF-QKD is given in Fig.~\ref{fig:TF_detection} and for MP-QKD in Fig.~\ref{fig:MP_detection}.

Considering the technology that is currently being developed for space applications ($Y_0 = 100$~Hz $ \Leftrightarrow  p_d = 4 \times 10^{-8}$, $\eta_D = 50 \%$~\cite{you2018superconducting}), we predict that a positive secret key rate can be obtained only for MP-QKD with $D_{\text{OGS}} = 100$~cm, with a maximal key rate of $R = 6.82 \times 10^{-9}$~bit/pulse, corresponding to 17 bit/s at the nominal repetition rate. This illustrates the limit of feasibility of GEO QKD exchange with current single-photon detection devices and small OGS telescope diameters and the importance of the evolution of the detector technology. Indeed, considering a space detector with characteristics that match those of ground-based technology, it would be possible to reach key rates of 280 bit/s and 822 bit/s for MP-QKD and TF-QKD respectively with a 1m OGS, and a positive key rate even for a 20~cm OGS diameter with both protocols. It is worth highlighting that TF-QKD is more sensitive to detector performance with respect to MP-QKD, as it can outperform the latter protocol only for better detector characteristics.

\section{Conclusion}
In this work, we have demonstrated the feasibility of a global-scale QKD link via a single GEO satellite equipped with two 50~cm telescopes, communicating with terrestrial optical ground stations with apertures ranging from 20~cm to 1~m. Two key elements are essential for achieving such a practical and compact system, namely the use of advanced QKD protocols that are exceptionally resilient to channel losses and the implementation of optimized AO beam pre-compensation in the uplink channels.
To derive the expected secret key rates, we developed a full end-to-end  channel model, considering atmospheric effects and AO pre-compensation of the optical beams. This model allowed us to assess the performance of two MDI-QKD-type protocols under such conditions: TF-QKD and MP-QKD, thus setting the limits of these protocols with current and future technology on detection, emission and turbulence mitigation. 
The results showed that considering the state-of-the-art detection systems for space applications, it would be possible to reach 17~bit/s with the MP-QKD protocol and two OGS of 1~m diameter. 
Considering an evolution of the detection system with performances close to those of ground-based solutions, the key rate for 1~m OGS would increase to 280~bit/s for MP-QKD and 822~bit/s for TF-QKD. Moreover, in such a scenario, it would be possible to obtain a positive key rate with OGSs down to 20~cm in diameter, highlighting the strong potential for scalability toward a global quantum communication network. This work offers a in-depth analysis of the feasibility of QKD protocols at a global scale, thus supporting the design of emerging global quantum communication networks. 
Moreover, our work highlights the impact of advanced AO pre-compensation methods, such as MMSE, in improving the link performances and eventually increasing the achievable key rate.
Future work will focus on extending the OGS network with LEO/GEO satellites and optimizing network performance through characterization of channel asymmetries.

\begin{acknowledgments}
We acknowledge financial support from the European Union’s Horizon Europe research and innovation programme under the project QSNP (Grant No. 101114043) and the project QUDICE (Grant No. 101082596), and the PEPR integrated project QCommTestbed (ANR-22-PETQ-0011), part of Plan France 2030.
\end{acknowledgments}

% --- REFERENCES ---
\bibliographystyle{apsrev4-2}
\bibliography{bib.bib}

\appendix

\section{Atmospheric turbulence induced losses model}
\label{appendix:atm_turb}
To simulate the turbulence impact on the optical link, we use a pseudo-analytical model - pseudo-analytical as we consider the phase and amplitude spatial statistics after propagation, but still rely on a numerical final step to compute the coupling losses induced by the phase distortions, as there is no model in the literature yet to directly describe this loss term statistics. Using a pseudo-analytical model has the great benefit of heavily reducing the simulation computation complexity, by suppressing the optical propagation step which is computationally intensive in end-to-end wave optics simulations. Hence, this model allows generating large datasets of uncorrelated turbulence loss samples.

\subsection{Reciprocal uplink losses}
To model the uplink turbulence-induced losses, we adopt a reciprocal formalism. The reciprocity principle states that the coupling efficiency of an emitted mode, propagated and coupled to a receiver mode, is equal to the coupling efficiency of this receiver mode back-propagated to the emitter and coupled to the emission mode. This principle is valid for turbulent medium and AO corrected links, as long as the medium is invariant within the propagation time of the beam through this medium, and if the correction is the same for both the up and downlink beams. 
Applying this principle, we can rewrite the coupled flux onboard the satellite as:
\begin{align}
    \eta_{\text{turb}} &= \eta_{\text{pre-compensated},\text{OGS}\rightarrow \text{Sat}}\\
    &= \eta_{\text{compensated},\text{Sat}\rightarrow \text{OGS}}.
\end{align}
This principle allows to model the uplink as a downlink at point-ahead angle, which enables us to use downlink modeling tools from the literature.
The turbulent losses can therefore be modeled as the following overlap integral:
\begin{equation}
\begin{split}
    \eta_{\text{turb}} &=\frac{\left| \iint_P \Psi_{\text{corr}}(\mathbf{r};\alpha_{\text{PAA}})M_0(\mathbf{r})d^2r\right|^2}{\iint_P\left|M_0(\mathbf{r})\right|^2d^2\mathbf{r}},
\end{split}
\end{equation}
where $\Psi_{\text{corr}}(\mathbf{r};\alpha_{\text{PAA}})=Ae^{\chi(\mathbf{r;\alpha_{\text{PAA}}})+j\Phi_{\text{corr}}(\mathbf{r};\alpha_{\text{PAA}})}$ is the downlink complex field at PAA corrected by adaptive optics, where $A$ is a constant amplitude term, $\chi(\mathbf{r};\alpha_{\text{PAA}})$ depicts the logarithm of the amplitude fluctuations (log-amplitude), $\Phi_{\text{corr}}(\mathbf{r};\alpha_{\text{PAA}})$ is the phase at PAA after AO correction, and $M_0(\mathbf{r})$ is the transmitter mode.

To model the losses of the reciprocal uplink corrected by adaptive optics, we assume the phase and log-amplitude contributions to the coupling losses to be independent, as follows:
\begin{equation}
    \eta_{\text{turb}} = \rho_\Phi\rho_\chi
\end{equation}

In this suboptimal phase correction scenario, there is no statistical law to describe $\eta_{\text{turb}}$. 
However, the statistics of the phase and log-amplitude of $\Psi_{\text{corr}}(\mathbf{r};\alpha_{\text{PAA}})$ is known from the literature~\cite{Mahe2000,Chassat_1989}. 
This allows us to use a pseudo-analytical model to generate an empirical statistical distribution of $\eta_{\text{turb}}$. We call this pseudo-analytical as we use the knowledge of the known phase and log-amplitude statistics to draw phase and log-amplitude samples, used to synthesize the complex field and perform numerically the overlap integral to the Gaussian mode of the transmitter. 
An in-depth description of the pseudo-analytical approach used in our analysis can be found in~\cite{lognone2023phase,lognone2022channel}.

\subsection{Log-amplitude induced losses}
\begin{figure}[b]
    \centering
    \includegraphics[width=1\linewidth]{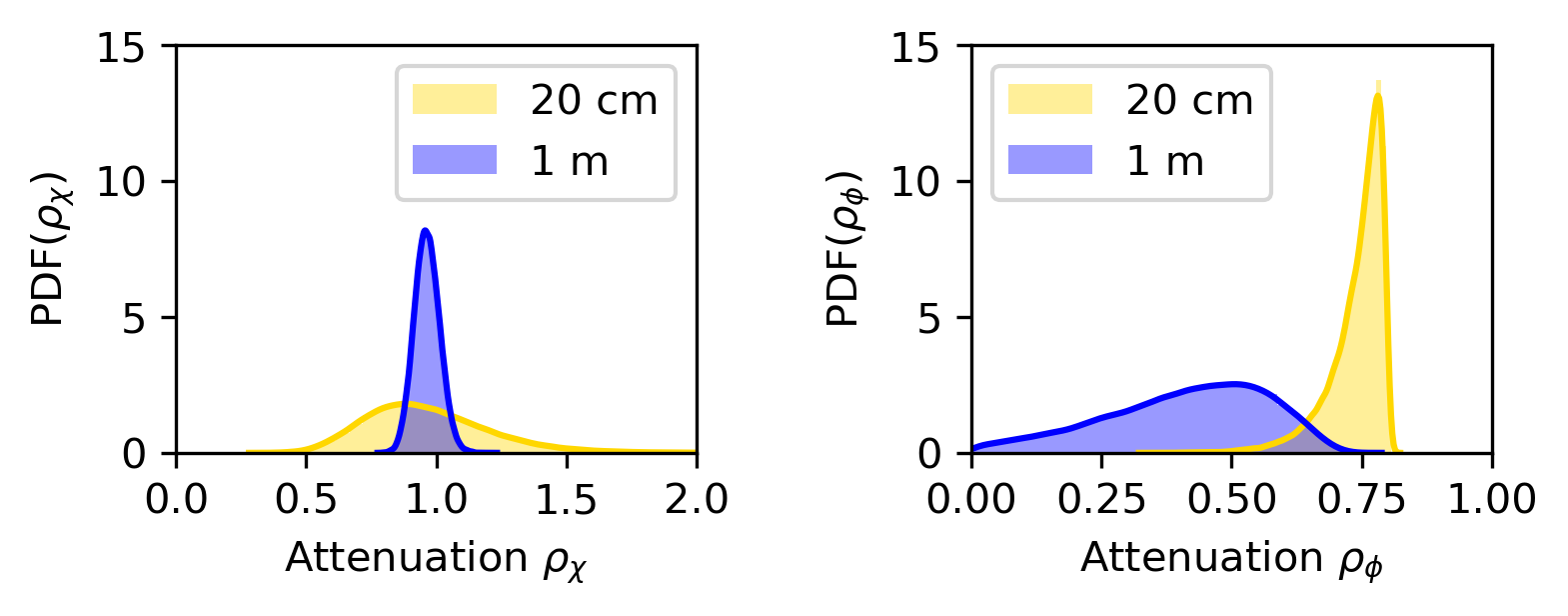}
    \caption{On the left, distribution of the log-amplitude induced fluctuations for the diameters 20~cm (yellow) and 1~m (blue). On the right, distribution of the phase induced attenuation for the diameters 20~cm (yellow) and 1~m (blue).}
    \label{fig:phase_amp_contribution}
\end{figure}
We assume the aperture averaged scintillation to dominate the log-amplitude contribution $\rho_\chi$. Therefore, $\rho_\chi$ is expressed as~\cite{Canuet:18,lognone2022channel}:
\begin{equation}\label{chi_contrib}
    \rho_\chi = e^{-\sigma^2_\chi}e^{-2\chi_{Ap}},
\end{equation}
where $e^{-\sigma^2_\chi}$ is a static penalty term to account for the spatial log-amplitude fluctuations~\cite{Mahe2000,sasiela}, and $\sigma_\chi^2$ denotes the log-amplitude variance defined as:
\begin{equation}
     \sigma_{\chi}^2=0.5631 k_0^{7/6}\int_0^Ldz C_n^2(z) z^{5/6} ,
\end{equation}
where $k_0$ is the wave number and $C_n^2(z)$ is the turbulence profile at distance $z$ from the OGS aperture. Additionally, $\chi_{Ap}$ is the log-amplitude averaged by the aperture random variable that follows a Normal distribution $\mathcal{N}(-\sigma^2_{\chi_{Ap}},\sigma^2_{\chi_{Ap}})$, whose variance is computed as~\cite{fried1967aperture,Mahe2000}:
\begin{equation}
\begin{split}
    \sigma^2_{\chi_{Ap}} = 5.20&R_{tel}^{5/3}k_0^2 \int_0^L dz C_n^2(z) \cdot\\
    &\int_0^\infty dk k^{-14/3} J_1(k)^2 \text{sin}^2\left(\frac{zk^2}{2k_0^2R_{\text{tel}}^2}\right),
\end{split}
\end{equation}
where $R_{\text{tel}}$ is the telescope aperture radius and $J_1$ is the Bessel function of the first kind of order~1.
Hence, we can draw an arbitrarily large number of random occurrences of $\chi_{Ap}$ and compute $\rho_\chi$. 
The distribution of $\rho_\chi$ is illustrated for two different apertures on the left of Fig.~\ref{fig:phase_amp_contribution} for two different aperture sizes. We observe large fluctuations of the log-amplitude losses for the 20~cm aperture diameter, and smaller fluctuations for the 1~m case, illustrating the aperture averaging property of large apertures. Additionally, the distribution of $\rho_\chi$ is centered around one. This is explained by the fact that the beam spatial amplitude fluctuations induced by atmospheric turbulence will either concentrate or dilute the power received in the aperture. 

\subsection{Turbulent phase induced losses}
\label{appendix:turb_model}
\subsubsection{General expression}
The phase contribution to the coupling $\rho_\Phi$ is derived as the overlap integral of the complex field, neglecting the log-amplitude fluctuation, to the Gaussian mode $M_0(\mathbf{r})$, therefore expressed as:
\begin{equation}
\label{eq_appendix:phase_contrib}
    \rho_\Phi = \exp(-\sigma^2_{\text{super-fitting}})\iint e^{j\Phi_{\text{res}}(\mathbf{r})}M_0(\mathbf{r})P(\mathbf{r})d\mathbf{r}^2 ,
\end{equation}
where $\exp(-\sigma^2_{\text{super-fitting}})$ is a constant loss induced by the unmodeled phase, where $\sigma^2_{\text{super-fitting}}=0.458(n_{r,\text{max}}+1)^{-5/3}\left(\frac{D}{r_0}\right)^{5/3}$~\cite{conan1994}, $\Phi_{\text{res}}(\mathbf{r})$ is the pre-compensation phase error, $M_0(\mathbf{r})$ is the Gaussian mode of the transmitter laser, of waist $\omega_0 = D/2.2$-\cite{Canuet:18}, and $P(\mathbf{r})$ is the circular mask of the telescope aperture. The spatial coordinates $\mathbf{r}$ are expressed in the aperture plane of the telescope.

\subsubsection{Spatial phase correction and associated statistics}
In adaptive optics systems, multiple errors affect the phase correction, namely, the temporal error - induced by the AO loop delay, the fitting error - induced by the limited number of correction modes of the AO system, the aliasing error, and, in the uplink case, the anisoplanatic error - induced by direction difference between the measured phase and the corrected phase. In this scenario, we assume the uplink pre-compensation phase error to be dominated by the fitting and anisoplanatic errors.
Therefore, the pre-compensation phase error is defined as $\Phi_{\text{res}}(\mathbf{r}) = \Phi_{\text{PAA}}(\mathbf{r}) - \Phi_{\text{AO}}(\mathbf{r})$, where $\Phi_{\text{PAA}}(\mathbf{r})$ is the phase at PAA we intend to correct, and $\Phi_{\text{AO}}(\mathbf{r})$ is the AO correction phase.

The spatial phase $\Phi(\mathbf{r})$ can be expressed over the telescope circular aperture as a vector of projection onto the Zernike polynomial basis, which is an orthogonal basis;
\begin{equation}
\mathbf{\Phi} = \begin{bmatrix} a_2, ... , a_N,
\end{bmatrix}
\end{equation}
where
\begin{equation}
a_i = \iint \Phi(\mathbf{r})Z_i(\mathbf{r})P(\mathbf{r})d\mathbf{r},
\end{equation}
where $P(\mathbf{r})$ is the telescope circular aperture mask and $Z_i(\mathbf{r})$ is the $i^{th}$ Zernike mode. 

In this formalism, the phase vectors are known to be random vectors following the Gaussian law $\mathcal{N}(\mathbf{0},\mathbf{\Gamma}_{\Phi_{\text{res}}})$~\cite{roddier_1999}, where the covariance matrix $\mathbf{\Gamma}_{\Phi_{\text{res}}}$ is expressed as:
\begin{equation}
    \mathbf{\Gamma}_{\Phi_{\text{res}}} = \begin{bmatrix}
        [\mathbf{\Gamma}_{\Phi_{\text{res,AO}}}]_{2\leq i,j \leq N_{\text{AO}}} & \boldsymbol{0} \\
        \boldsymbol{0} & [\mathbf{\Gamma}_{\Phi\Phi}(0)]_{N_{\text{AO}}+1\leq i,j \leq N_{\text{max}}}
    \end{bmatrix},
\end{equation}
where $\mathbf{\Gamma}_{\Phi_{\text{res,AO}}}$ is the covariance of the phase corrected by the AO system (from mode $2$ to mode $N_{\text{AO}}$) and $\mathbf{\Gamma}_{\Phi\Phi}(0)$ is the autocovariance of the turbulent phase, that is uncorrected by the AO system (from mode $N_{\text{AO}}+1$ to $N_{\text{max}}$, the maximum mode used in the representation).

In this study, we consider two types of correction: the state-of-the-art (SoA) and the MMSE correction. The classical correction consists in correcting the phase at PAA with the on-axis phase, and is expressed as:
\begin{equation}
    \mathbf{\Phi}_{\text{res,SoA}} = \mathbf{\Phi}_{\text{PAA}} - \mathbf{\Phi}_{0}
\end{equation}
In this case, the residual phase covariance matrix is expressed as:
\begin{equation}
    \mathbf{\Gamma}_{\Phi_{\text{res,SoA}}} = 2\mathbf{\Gamma}_{\Phi\Phi}(0) - \mathbf{\Gamma}_{\Phi\Phi}(\alpha) - \mathbf{\Gamma}_{\Phi\Phi}^T(\alpha),
\end{equation}
where $\mathbf{\Gamma}_{\Phi\Phi}(0)$ is the phase autocovariance matrix and $\mathbf{\Gamma}_{\Phi\Phi}(\alpha)$ is the phase angular covariance matrix. 

In the MMSE correction case, the phase at point-ahead angle is estimated using an MMSE estimation, performed on the downlink phase and amplitude measurements, and relying on the knowledge of the phase and amplitude angular statistics. In this case, the residual phase is expressed as:
\begin{equation}
    \mathbf{\Phi}_{\text{res,MMSE}} =  \mathbf{\Phi}_{\text{PAA}} - \mathbf{R}_{\text{MMSE}}\mathbf{y}_m,
\end{equation}
where $\mathbf{R}_{\text{MMSE}}$ is the phase reconstructor and ${y}_m$ is the downlink measurement vector, composed of the downlink phase and amplitude. In this case, the residual phase covariance matrix is expressed as:
\begin{equation}
    \mathbf{\Gamma}_{\Phi_{\text{res,MMSE}}} = \mathbf{\Gamma}_{\Phi\Phi}(0) - \mathbf{R}_{\text{MMSE}} \mathbf{\Gamma}_{\Phi y_m}^T(\alpha),
\end{equation}
where $\mathbf{\Gamma}_{\Phi y_m}(\alpha)$ is the angular covariance matrix between the phase at PAA and the measurement vector.

The complete expression of the reconstructor $\mathbf{R}_{\text{MMSE}}$, along with the formulas to compute the content of the different covariance matrices can be found in~\cite{lognone2023phase}.

Knowing the phase statistics, random phase vectors can be drawn, synthesized to a spatial phase and numerically coupled to the Gaussian mode, following Eq.~(\ref{eq_appendix:phase_contrib}).

\section{MDI-QKD simulation model}
\subsection{Security model for asymmetric twin-field QKD}
\label{sec:TF_security_model}
The model used was proposed in~\cite{wang2020simple}. To compensate for asymmetry, the protocol suggests adjusting the signal intensities such that the arriving intensities at Charlie's side are balanced, satisfying the condition:
\begin{equation}
\gamma_A = \alpha_A^2 \eta_A, \quad \gamma_B = \alpha_B^2 \eta_B.
\end{equation}
In our case, instead of adjusting the intensity of the pulses to get a symmetrical attenuation on both channels, we apply a compensation at Charlie's side to create the same conditions. This means that we do not need to simulate different intensities on each side. This adjustment minimizes the $X$-basis bit error rate, $e_{XX}$, which is directly impacted by channel asymmetry.

The gain in the $X$-basis is given by:
\begin{align}
\begin{split}
p_{XX} &= \frac{1}{2}(1 - p_d) [e^{-\sqrt{\gamma_A \gamma_B} \cos(\theta) \cos(\phi)} \\
& + e^{\sqrt{\gamma_A \gamma_B} \cos(\theta) \cos(\phi)} ] e^{-\frac{1}{2}(\gamma_A + \gamma_B)} \\
&- (1 - p_d)^2e^{-(\gamma_A + \gamma_B)},
\end{split}
\end{align}
where the detector dark count probability is $p_d$, the polarization misalignment between Alice and Bob $\theta$, and the phase mismatch between Alice and Bob $\phi$. The $X$-basis bit error rate is:
\begin{align}
\begin{split}
    e_{XX}&= (e^{-\sqrt{\gamma_A \gamma_B} \cos(\theta) \cos(\phi)} - (1-p_d)e^{-\frac{1}{2}(\gamma_A + \gamma_B)})\\
    & \times (e^{-\sqrt{\gamma_A \gamma_B} \cos(\theta) \cos(\phi)} + e^{\sqrt{\gamma_A \gamma_B} \cos(\theta) \cos(\phi)}\\ 
    &- 2(1-p_d)e^{-\frac{1}{2}(\gamma_A + \gamma_B)})^{-1}.
\end{split}
\end{align}

The gain in the $Z$-basis, which incorporates all possible relative phases between the signals, is expressed as:
\begin{align}
\begin{split}
p_{ZZ}& = (1-p_d) \cdot q_{ZZ}\\
& + (1-p_d)p_d(1-\eta_A)^{n_A}(1-\eta_B)^{n_B},
\end{split}
\end{align}
including the infinite-decoy case. The formula for $q_{ZZ}$ is described in~\cite{wang2020simple}. The security of the protocol is achieved by bounding the phase error rate, $e_{ZZ}$, which is obtained through a finite decoy-state analysis. This rate is upper-bounded as:
\begin{equation}
p_{XX} \cdot e_{ZZ} \leq \sum_{n, m} \sqrt{Y_{nm} \cdot p_{ZZ}},
\end{equation}
where $Y_{nm}$ represents the yield for $n$-photon (Alice) and $m$-photon (Bob) states. Using these bounds, the secret key rate is calculated as:
\begin{equation}
R = 2 \cdot p_{XX} \left[1 - f_{EC}  H(e_{XX}) - H(e_{ZZ})\right],
\end{equation}
where $H(x)$ is the binary entropy function and $f_{\text{EC}}$ is the error correction efficiency.

\subsection{Security model for asymmetric mode-pairing QKD}
\label{sec:MP_security_model}
The model used was proposed in~\cite{lu2024asymmetric}. The key rate \(R\), in the asymptotic case, is expressed as:
\begin{equation}
R = r_p(p, L_{\text{max}}) r_s \left[ q_{(1,1)} \left( 1 - H(e_{(1,1)}) \right) - f_{\text{EC}} H(e_{Z}) \right],
\end{equation}
where $r_p(p, L_{\text{max}})$ is the pairing rate with $p$ the successful click probability in each round and $L_{\text{max}}$ the maximal pairing length, \(r_s\) is the \(Z\)-pair ratio, \(q_{(1,1)}\) is the single-photon pair ratio, \(e_{(1,1)}\) is the phase error rate, and \(H(x)\) is the binary entropy function. The pairing rate, which reflects the probability of forming valid pulse pairs, varies with the maximum pairing interval $L_{\text{max}}$. The $Z$-pair ratio $r_s$ and the pairing ratio $r_p$ are given by:
\begin{equation}
    r_s = \frac{1}{16p^2}\sum \limits_{z_i \oplus z_j = 11} \Pr(C_i = 1 \mid z_i)\Pr(C_j = 1 \mid z_j),
\end{equation}
\begin{equation}
    r_p = 
\left[
\frac{1}{p[\left[(1 - p)^{L_{\text{min}} - 1} - (1 - p)^{L_{\text{max}}}\right]} + \frac{1}{p}
\right]^{-1},
\end{equation}
where $L_{\text{min}}$ is the minimal pairing length introduced to account for the detector dead time. Here, the decoy-state estimation is analyzed with the infinite key size. Decoy-state analysis allows reenforcing the security of the protocol by bounding the single-photon pair ratio and the phase error rate. The single-photon pair ratio \(q_{(1,1)}\) is given by:
\begin{align}
\begin{split}
q_{(1,1)} = & \frac{1}{16} \frac{P_{\mu^a}(1) P_{\mu^b}(1)}{r_s p^2}  [  \sum_{z_i \oplus z_j = 11} \Pr(C_i = 1 \mid n_i = z_i) \\
 & \times \Pr(C_j = 1 \mid n_j = z_j) ],
\end{split}
\end{align}
where \(\eta_A\) and \(\eta_B\) are the channel transmittances for Alice and Bob, and $P_{\mu^{a(b)}}(k)$ is the Poisson distribution. The $X$-basis gain $Y_{(1,1)}$ and phase error rate \(e_{(1,1)}\) are determined using:
\begin{align}
\begin{split}
Y_{(1,1)} = &(1 - p_d)^2 
[\frac{\eta_A \eta_B}{2} 
+ \left(2\eta_A + 2\eta_B - 3\eta_A \eta_B\right)p_d \\
&+ 4(1 - \eta_A)(1 - \eta_B)p_d^2] ,\\
e_{(1,1)} = &\frac{e_0 Y_{(1,1)} - (e_0 - e_d)(1 - p_d^2)\frac{\eta_A \eta_B}{2}}{Y_{(1,1)}}.
\end{split}
\end{align}
Here, \(e_0 = 0.5\) accounts for errors caused by vacuum pulses, and \(e_d\) refers to misalignment errors. This error rate is directly responsible for the limitation of the maximal pairing length in the MP-QKD scheme.

\section{Phase fluctuation model}
\label{sec:phase_fluct}
The phase fluctuation model for MP-QKD used is similar to the model introduced ~in~\cite{zhu2023experimental} and \cite{zhou2025optimization}. The phase difference between the pulses sent by Alice and Bob is mainly due to two phenomena: laser imperfection and free-space fluctuations. The phase difference for two incoming pulses from Alice and Bob can be expressed as:
\begin{equation}
    \theta_{ba} = \theta_{ba}^0 + (\theta_{\text{fs},b} - \theta_{\text{fs},a}) + (\nu_b - \nu_a)t,
\end{equation}
where \(\nu_a\) and \(\nu_b\) are the angular frequencies of the light pulses, and \(t\) is the transmission time. The initial phase difference between Alice’s and Bob’s lasers is \(\theta_{ba}^0\). With system synchronization, the transmission times of Alice’s and Bob’s pulses are identical and only depend on the pairing interval achieved. Finally, $\theta_{\text{fs},b}$ and $\theta_{\text{fs},a}$ are the phase drifts after the free-space propagation for each channel.
In MP-QKD, we are only interested in the phase differences of the paired $i$-th and $j$-th rounds so the phase difference between both sides becomes:
\begin{equation}
\label{eq:phase}
\Delta \theta_{i,j} = \Delta \theta_{ba}^0 + (\Delta \theta_{\text{fs},b} - \Delta \theta_{\text{fs},a}) + \Delta \nu(t_j - t_i),
\end{equation}
where \(t_i\) and \(t_j\) are the transmission times of the \(i\)-th and \(j\)-th rounds, respectively. The additional phase differences induced by the free-space channel between the \(i\)-th and \(j\)-th rounds from each side are represented by \(\Delta\theta_{\text{fs},a}\) and \(\Delta\theta_{\text{fs},b}\). The difference in initial phase is denoted as \(\Delta\theta_{ba}^0\) and represents the linewidth impact. The time difference between the $i$-th and $j$-th rounds can be expressed as $t_j - t_i = \frac{L}{F}$ with $F$ being the repetition rate. A more precise description of these effects is given in the next paragraph.\\

\begin{figure}[h]
    \centering
    \includegraphics[width=1\linewidth]{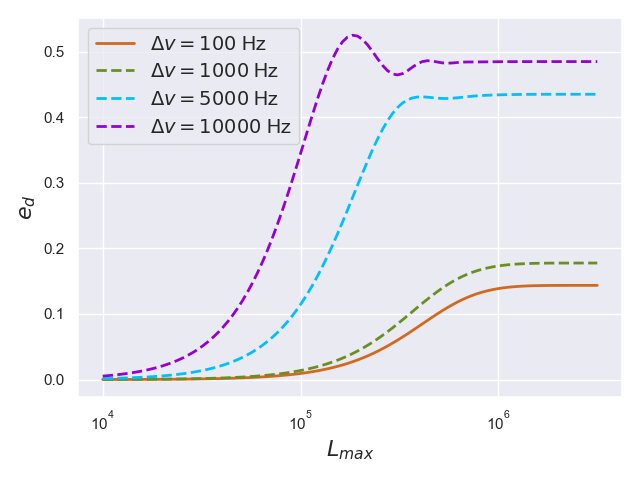}
    \caption{MP-QKD $X$-basis misalignment error evolution with $L_{\text{max}}$.}
    \label{fig:phase_evolution}
\end{figure}

\textbf{Evolution of the phase difference:}
Using the same method as in~\cite{zhou2025optimization}, we consider that the frequency
uncertainty caused by linewidth follows a Gaussian distribution centered around its central frequency and with a standard deviation denoted $\sigma_{\nu}$. Therefore, for two different rounds, the linewidth effect follows a Gaussian distribution with a standard deviation of $\sqrt{2}\sigma_\nu$. Then, we approximate the distribution of the free-space phase drift from Fig.~\ref{fig:histo} as a centered Gaussian with standard deviation of $\sigma_{\text{fs}} = 150$ rad/s. The $X$-basis phase error for one pair is: 
\begin{equation}
\begin{split}
    \text{e}_{\text{ph}}(L) &=
    \int_{-\infty}^{\infty} \int_{-\infty}^{\infty} 
     \frac{1 - \cos \Delta \theta}{2} G(\nu) G(\omega_{\text{fs}}) \, d\nu \, d\omega_{\text{fs}}\\
    &= \frac{1}{2} - \frac{1}{2}
    \int_{-\infty}^{\infty} \int_{-\infty}^{\infty} \cos \left[(2\pi \Delta v + 2\pi v + \omega_{\text{fs}}) \Delta t \right]\\
    & \qquad  \cdot G(\omega_{\text{fs}}) G(\nu)  \, d\nu \, d\omega_{\text{fs}}\\
    & =\frac{1}{2} - \frac{1}{2} e^{-\sigma^2_{\text{fs}} \Delta t^2 / 2} e^{-2\sigma^2_v (2\pi \Delta t)^2 / 2} \cos(2\pi \Delta v \Delta t)\\
    &= \frac{1}{2} - \frac{1}{2} e^{-\sigma^2_{\text{tot}} \left(\text{L}/{F}\right)^2 / 2} \cos\left(2\pi \Delta v \cdot \frac{\text{L}}{F}\right),
\end{split}
\end{equation}
with $\sigma^2_{\text{tot}} = \sigma^2_{\text{fs}} + 8\pi^2\sigma_{\nu}^2$. 
\begin{figure}[t]
    \centering
    \includegraphics[width=0.45\textwidth]{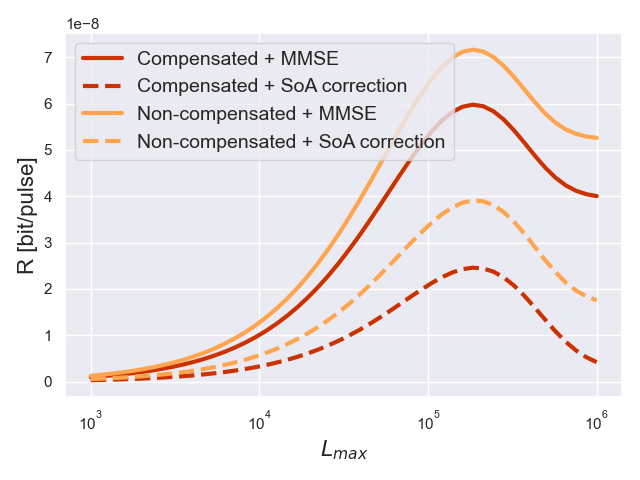}
    \caption{MP-QKD key rate performance with respect to $L_{\text{max}}$. $D_{OGS} = 100$ cm, $\mu = 0.60$, $p_d = 10^{-8}$, $\eta_D = 70 \%$, $L_{\text{min}} = 100$.}
    \label{fig:mp_perf}
\end{figure}
Furthermore, we also need to provide a distribution of $L$ as a function of $L_{\text{max}}$ and $L_{\text{min}}$ to account for the time period between the two pulses in a successful round. Since we only want to know the phase fluctuation assuming that we have a successful round, the probability of a round consisting of $n$ pulses, if we have a pair created, is:
\begin{equation}
    P(L = n) = \frac{ p \cdot (1-p)^{n-1}}{1 -  (1-p)^{L_{\text{max}}}} .
\end{equation}
Then, $e_d$ can be regarded as the weighted average of $e_{\text{ph}}(L)$ with $L$ ranging from $L_{\text{min}}$ to $L_{\text{max}}$. The $X$-basis phase error is therefore:
\begin{equation}
\label{eq:ed}
    e_d = \sum_{n = L_{\text{min}}}^{L_{\text{max}}} P(L = n) \cdot e_{\text{ph}}(n) .
\end{equation}

In Fig.~\ref{fig:phase_evolution} we show the evolution of the misalignment error $e_d$ with $L_{\text{max}}$ for several frequency offsets. We chose to work with $\Delta \nu = 0.1$ kHz in this analysis.

\section{MP-QKD performance evaluation with $L_{\text{max}}$}
\label{sec:qkd_simu}

We must compare the MP-QKD performances for different values of maximal pairing lengths, since $L_{\text{max}}$ has a non-negligible impact on the $X$-basis error rate and on the key rate. The performance is shown in Fig.~\ref{fig:mp_perf}.

For this configuration, $L_{\text{max}} = 184206$ is the optimal maximal pairing length. Moreover, according to Eq.~(\ref{eq:ed}), the performance is highly dependent on the probability of detection $p$. Thus, if $p$ is good enough, increasing the maximal pairing length after a certain threshold would not impact the performance since the probability of a round consisting of $n$ pulses becomes approximately constant, allowing to pair the two pulses no matter the chosen $L_{\text{max}}$.

\end{document}